\documentclass[prb,showpacs,amsmath,amssymb,twocolumn]{revtex4}

\usepackage{graphicx}
\usepackage{epsfig}
\usepackage{dcolumn}
\usepackage{bm}

\begin{document}

\title{Quantum dots in Si/SiGe 2DEGs with Schottky top-gated leads}

\author{K~A~Slinker$^1$, K~L~M~Lewis$^1$, C~C~Haselby$^1$, S~Goswami$^1$, L~J~Klein$^1$, 
J~O~Chu$^2$, S~N~Coppersmith$^1$, Robert Joynt$^1$, R~H~Blick$^1$, Mark Friesen$^1$ and 
Mark A~Eriksson$^1$}
\affiliation{
$^1$University of Wisconsin-Madison, Madison, WI 53706, \\
$^2$IBM Research Division, T.~J.~Watson Research Center, NY 10598}

\begin{abstract}
We report on the fabrication and characterization of quantum dot devices in a Schottky-gated silicon/silicon-germanium two-dimensional electron gas (2DEG).  The dots are confined laterally inside an etch-defined channel, while their potential is modulated by an etch-defined 2DEG gate in the plane of the dot.  For the first time in this material, Schottky top gates are used to define and tune the tunnel barriers of the dot.  The leakage current from the gates is reduced by minimizing their active area.  Further suppression of the leakage is achieved by increasing the etch depth of the channel. The top gates are used to put the dot into the Coulomb blockade regime, and conductance oscillations are observed as the voltage on the side gate is varied.
\end{abstract}


\maketitle

A new generation of high quality, high mobility silicon/silicon-germanium quantum well devices has recently emerged to meet the needs of nascent technologies like quantum computing and spintronics.  In the near future, it is possible that silicon will form the basis for many of the same nano-scale devices previously constructed from III-V materials.  Indeed, silicon has some advantages over other materials in the context of quantum information processing due to long spin lifetimes associated with low spin-orbit 
coupling.\cite{Tyryshkin,Wilamowski,Gordon,Feher,Feher2,Chibi}  Quantum computing architectures have been proposed, requiring not only the unprecedented control of individual few-electron quantum dots, but also the fine-tuning of couplings to neighboring 
dots.\cite{Loss,Vrijen,Friesen,Friesen2,Eriksson}

To meet these needs, an expanding fabrication toolbox has been developed for silicon/silicon-germanium quantum well devices.  Etch-defined quantum dots have been fabricated in Si/SiGe quantum wells\cite{Klein} and bulk SiGe\cite{Qin,Bo} using lateral side gates in the plane of the dot.  These lateral gates can successfully modulate the potential of the dot.  However, the etching procedure induces a depletion zone around the etch boundary which limits the control of the tunnel barriers\cite{Klein2} due to the weak capacitive coupling of the side gates.  Metal side gates can be used to reduce the distance between the dot and the gates,\cite{Sakr} thereby improving the coupling.  Nevertheless, depletion zones cannot be avoided.  Since exquisite control of the tunnel barriers between dots and leads (or additional dots) is a prerequisite for quantum computing, quantum dot tunnel barriers are unlikely to rely on side gates for their tuning.  

In gallium arsenide quantum dots, a level of control consistent with quantum computing has been achieved using Schottky top gates.\cite{Elzerman}  In this technique, metal gates are patterned directly above the quantum well heterostructure, providing a strong and proximal capacitive coupling.  Current flow between the top gates and the underlying two-dimensional electron gas (2DEG) is prevented by the Schottky barrier, which depletes carriers in the vicinity of the electrodes.  Similar schemes have been explored in the context of silicon heterostructures, culminating in the demonstration of quantum point 
contacts.\cite{Koester,Tobben,Tobben2,Wieser}  However, silicon Schottky barriers have proven 
leaky,\cite{Dunford} with current paths possibly following lattice dislocations, arising from state-of-the-art strain relaxation techniques in Si/SiGe heterostructures.  The leakage problem, which forms the greatest obstacle for silicon quantum devices, is still poorly understood.  In short, Schottky top-gated quantum dots have not been attained in modulation-doped silicon/silicon-germanium 2DEGs.  Their demonstration forms an important milestone.  

Here, we demonstrate two hybrid quantum dot devices that employ both etching and top-gate fabrication techniques.  In particular, we show that tunnel junctions can be formed and tuned using Schottky top gates.  An additional degree of control is provided by a back gate, which enables the density of the 2DEG to be varied. Leakage is minimized in this geometry due to the small size of the top gates, whose active area is less than 0.2~$\mu \text{m}^2$.  Additional suppression of leakage is observed in devices with deeper channel etches.  We hypothesize that ultra small gates reduce the overlap of electrodes with localized leakage paths.  

The devices are patterned in a Si/SiGe heterostructure grown by ultrahigh vacuum chemical vapor deposition.  The two-dimensional electron gas sits at the top of a 80~\AA~strained silicon layer capped with 140~\AA~of Si$_{0.7}$Ge$_{0.3}$, 140~\AA~of P-doped 
Si$_{0.7}$Ge$_{0.3}$, and 35~\AA~of Si.  The strained silicon is grown on a relaxed 
Si$_{0.7}$Ge$_{0.3}$ virtual substrate obtained by grading from 
0\% to 30\% Ge over several microns.\cite{Ismail}  Strained silicon is necessary to achieve the correct band offsets for quantum confinement relative to the surrounding (relaxed) 
SiGe.\cite{Schaffler}  
The electron density of the 2DEG is measured to be 
4$\times 10^{11} \text{ cm}^{-2}$ with a mobility of $40,000 \text{ cm}^2$/Vs at 2~K.  Ohmic contacts are made to the 2DEG by Au/Sb evaporation and subsequent 400$^\circ$~C anneal.  
Gold is sputtered onto the entire back of the sample to form a back gate. 

\begin{figure}[t]
\centerline{\includegraphics[width=2.3in]{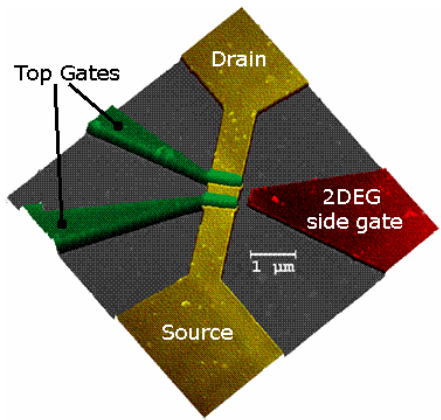}}
\caption{
False color atomic force micrograph of a quantum dot consisting of an etched-defined 2DEG channel, a 2DEG side gate in the same plane as the channel, and metal Schottky gates extending from the etched region up and across the channel.  A negative voltage is applied to the top gates to locally deplete the 2DEG and form a quantum dot between the gates.  The potential of the dot can be modulated by the side gate.
\label{fig:AFM}}
\end{figure}

A 4~$\mu$m long, 800~nm wide channel is defined in the heterostructure by e-beam lithography and CF$_4$ reactive ion etch (Fig.~\ref{fig:AFM}).  The etch depths are 55 and 85~nm for Devices A and B respectively, as confirmed by atomic force microscopy.  A single side gate about 300~nm from the channel is patterned into the heterostructure in the same lithographic and etching step.  The 2DEG in this lateral gate is electrically isolated from the 2DEG in the channel, and its capacitive coupling can be used to modulate the potential of the electrons in the quantum dot.

Metal top gates are formed by a second e-beam lithography step followed by evaporation of 
80~nm of Pd and liftoff.  The gates are patterned with an active top-gate area of 100~nm by 
800~nm and are carefully aligned to within 20~nm such that the gate just extends across the top of the channel.  Contact pads and wires are also formed in this step in the etched regions and extend up the side of the channel to the gates.  

\begin{figure}[b]
\centerline{\includegraphics[width=2.7in]{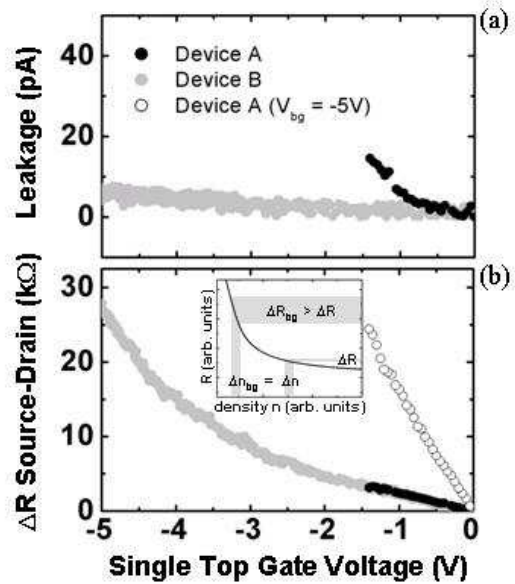}}
\caption{
(a) As the voltage is varied on a single top gate, the leakage current is minimal out to 
-1.4~V for Device A and -5~V for the deeper-etched Device B.  (b) The change in resistance through the channel from $V_\text{tg} = 0$~V within this low-leakage working range.  
For Device A, an increase in resistance greater than the $h/e^2$ necessary for quantum confinement is achieved only after a backgate is used to lower the carrier concentration in the channel.  Inset: The change in density as the top gate voltage is varied is the same with and without a negative voltage applied to the back gate.  However, the change in resistance 
($\Delta R_\text{bg}$) is greater at the lower initial density induced by the back gate than with no back gate voltage applied ($\Delta R$).
\label{fig:dR}}
\end{figure}

The Schottky gates are characterized by measuring the change in resistance through the channel of each device along with the leakage current from the gate as the voltage ($V_\text{tg}$) on a single top gate is varied (Fig.~\ref{fig:dR}).  The leakage current of Device A is small 
($< 15$~pA) down to an applied voltage of $-1.4$~V.  In this working range, the resistance through the channel increases from its value at $V_\text{tg} = 0$ by 6~k$\Omega$.  However the resistance through each lead should be greater than $h/e^2 = 26 \text{ k}\Omega$ when used to define the quantum dot.

Device B exhibits a much greater working range, with minimal leakage current ($< 10$~pA) out to $-5$~V –- more than 3 times greater than Device A -– such that a tunneling resistance of 
28~k$\Omega$ is achieved between the two sides of the channel.  The only difference in the fabrication of the two devices is the deeper etch depth for Device B.  Previous work suggests that sidewall depletion increases with etch depth in similarly etched Si/SiGe 2DEG 
wires\cite{Holzmann} -– a trend we have observed as well.  This wider depletion zone associated with the deeper etch may provide better insulation of the metal electrodes from the active 2DEG of the channel. 

In the case that the leakage current is still a limiting factor (as in Device A) the back gate can be employed to tune the carrier density in the device such that the necessary resistance through the leads is achieved within the working range of the top gates.  The carrier density decreases with a negative voltage applied to the back gate ($V_\text{bg}$) and the resistance of the whole device –- including the leads -– is increased.  Since we do not know the lead resistance separate from the rest of the device, we again measure the increase in resistance induced by a top gate.  Because the resistance is inversely proportional to the density, the resistance increases more rapidly given a lower carrier density in the channel as the voltage on the top gate is varied (inset of Fig.~\ref{fig:dR}).  The third data set in figure 2 shows that, with $V_\text{bg} = -5$~V, an increase in resistance of 25~k$\Omega$ is observed in Device A as the top-gate voltage is varied from 0~V to $-1.4$~V.  Voltages on the back gate down to $-7$~V are applied to achieve changes in resistance due to the top gate as high as 80~k$\Omega$. 

\begin{figure}[t]
\centerline{\includegraphics[width=2.7in]{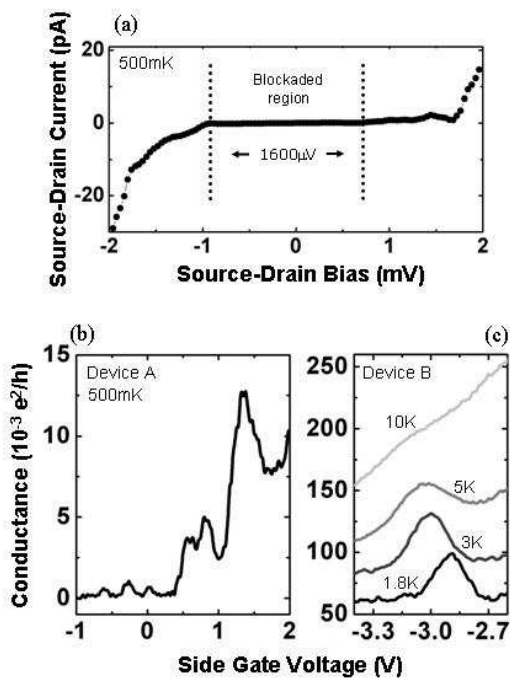}}
\caption{
(a) Current through the dot as the drain-source bias across the dot is varied.  The width of the zero-conductace plateau about $V_\text{ds} = 0$ corresponds to a charging energy of 
800~$\mu$V. (b) Conductance measurement through the dot as the voltage on the lateral 2DEG gate is varied.  Oscillations are observed indicating that the charge on the dot has changed by one electron between minima.  Two regimes of the dot are shown: 
(i) At $V_\text{sg} < 0.3$~V, the electrons in the dot are well isolated from the leads such that the conductance goes to zero between peaks.  
(ii) For $V_\text{sg} > 0.3$~V, the side gate has opened up the leads such that the dot is better coupled to the leads; the peaks are broadened and the conduction does not go to zero between peaks even when the background is removed.  (c) A single conductance peak as the temperature is varied.  The peak is indistinguishable from the background at 10~K, in good agreement with the measured charging energy of the dot.
\label{fig:IV}}
\end{figure}

The confining characteristics of the top gates are further demonstrated by putting both leads of a device into the tunneling regime.  A quantum dot is formed in the undepleted region between the gates, with side-wall confinement provided by the surface depletion from the etching.  Conductance through the channel is blockaded for source-drain biases ($V_\text{ds}$) in the range $-1\text{ mV} < V_\text{ds} < 0.6\text{ mV}$ as shown in Fig.~\ref{fig:IV}.  Assuming a disc geometry, this gives a dot diameter of 470~nm with about 700~electrons in the dot with the back gate grounded.  

The etch-defined lateral gate can be used to modulate the potential in the dot such that conductance oscillations are observed (Fig.~\ref{fig:IV}).  As the voltage on the side gate 
($V_\text{sg}$) is increased, the number of electrons in the dot increases by one with each oscillation.  However, the large separation between the dot and the side gate produces a non-proximal capacitive coupling, which also affects the tunnel barriers.  For 
$V_\text{sg} > 0.3$~V, the side gate voltage decreases the resistance in the leads such that finite background conductance is observed.  Conductance oscillations continue to be observed, however, indicating that electrons are still bound in the dot between the leads.  The conductance between the leads no longer goes to zero even with the background removed, suggesting that the tunnel coupling of the dot to the source and drain leads has 
increased.\cite{Kouwenhoven}  The peak spacing with Vsg is about 280~mV, giving a capacitive coupling of 0.57~aF that is similar to the coupling observed for lateral 2DEG gates used to modulate purely etch-defined dots.\cite{Klein}  These oscillations are observable up to about 10~K or 830~$\mu$eV (Fig.~\ref{fig:IV}), consistent with the charging energy calculated from the Coulomb blockade. 

In conclusion, we have shown that low-leakage Schottky top gates can be implemented in Si/SiGe heterostructures over a large working range in voltage, producing tunnel barriers and quantum dots in the underlying 2DEG.  In one device, an additional back gate was needed to achieve tunneling behavior within the working range of the top gates.  We anticipate that such back gates will become common features of silicon quantum devices, in analogy with III-V devices.  In a second device with deeper etching, a much larger working range was observed, increasing the resistance into the tunneling regime even without a back gate.  In future work, we believe that optimization of etch depths together with sidewall passivation or gate oxides could achieve better isolation results than either process alone.  Quantum dots were achieved in both our devices, with tunable Schottky barriers.  In addition to Schottky gates, the dots also utilize 2DEG side gates, demonstrating the compatibility of the two techniques.  For proximal control of individual dots, we expect that Schottky plunger gates will replace side gates, allowing for much smaller dots as well as multiple coupled dots.

This work was supported in part by the NSA and ARDA under ARO contract number W911NF-04-1-0389, and by NSF under Grant No. DMR-0325634 and DMR-0079983.



\begin{thebibliography}{99}

\bibitem{Tyryshkin}
Tyryshkin A M, Lyon S A, Astashkin A V and Raitsimring A M 2003 
\textit{Physical Review B (Condensed Matter and Materials Physics)} \textbf{68} 193207

\bibitem{Wilamowski}
Wilamowski Z and Jantsch W 2002 \textit{Physica E: Low-dimensional Systems and Nanostructures} 
\textbf{12} 439

\bibitem{Gordon}
Gordon J P and Bowers K D 1958 \textit{Phys. Rev. Lett.} \textbf{1} 368

\bibitem{Feher}
Feher G 1959 \textit{Phys. Rev.} \textbf{114} 1219-44

\bibitem{Feher2}
Feher G and Gere E A 1959 \textit{Phys. Rev.} \textbf{114} 1245-56

\bibitem{Chibi}
Chibi M and Hirai J 1972 \textit{Journal of the Physical Society of Japan} \textbf{33} 730

\bibitem{Loss}
Loss D and DiVincenzo D P 1998 \textit{Physical Review A (Atomic, Molecular, 
and Optical Physics)} \textbf{57} 120

\bibitem{Vrijen}
Vrijen R, Yablonovitch E, Wang K, Jiang H W, Balandin A, Roychowdhury V, Mor T and 
DiVincenzo D 2000 \textit{Phys. Rev. A} \textbf{62} 12306

\bibitem{Friesen}
Friesen M, Joynt R and Eriksson M A 2002 \textit{Applied Physics Letters} \textbf{81} 4619

\bibitem{Friesen2}
Friesen M, Rugheimer P, Savage D E, Lagally M G, van der Weide D W, Joynt R and 
Eriksson M A 2003 \textit{Physical Review B (Condensed Matter and Materials Physics)} 
\textbf{67} 121301

\bibitem{Eriksson}
Eriksson M A, et al. 2004 \textit{Quantum Information Processing} \textbf{3} 1

\bibitem{Klein}
Klein L J, et al. 2004 \textit{Applied Physics Letters} \textbf{84} 4047

\bibitem{Qin}
Qin H, Yasin S and Williams D A 2003 \textit{Journal Of Vacuum Science \& Technology B} 
\textbf{21} 2852-55

\bibitem{Bo}
Bo X Z, Rokhinson L P, Yin H Z, Tsui D C and Sturm J C 2002 \textit{Applied Physics Letters} 
\textbf{81} 3263-65

\bibitem{Klein2}
Klein L J, et al. Preprint cond-mat/0503766

\bibitem{Sakr}
Sakr M R, Yablonovitch E, Croke E T and Jiang H W Preprint cond-mat/0504046

\bibitem{Elzerman}
Elzerman J M, Hanson R, van Beveren L H W, Witkamp B, Vandersypen L M K and Kouwenhoven L P 2004 \textit{Nature} \textbf{430} 431-35

\bibitem{Koester}
Koester S J, Ismail K, Lee K Y and Chu J O 1997 \textit{Applied Physics Letters} 
\textbf{71} 1528

\bibitem{Tobben}
Tobben D, Wharam D A, Abstreiter G, Kotthaus J P and Schaffler F 1995 
\textit{Physical Review B (Condensed Matter)} \textbf{52} 4704

\bibitem{Tobben2}
Tobben D, Wharam D A, Abstreiter G, Kolthaus J P and Schaffler F 1995 
\textit{Semiconductor Science and Technology} 711

\bibitem{Wieser}
Wieser U, Kunze U, Ismail K and Chu J O 2002 
\textit{Physica E: Low-dimensional Systems and Nanostructures} \textbf{13} 1047

\bibitem{Dunford}
Dunford R B, Griffin N, Paul D J, Pepper M, Robbins D J, Churchill A C and Leong W Y 2000 
\textit{Thin Solid Films} \textbf{369} 316

\bibitem{Ismail}
Ismail K, Arafa M, Stern F, Chu J O and Meyerson B S 1995 \textit{Applied Physics Letters} 
\textbf{66} 842

\bibitem{Schaffler}
Sch\"{a}ffler F 1997 \textit{Semiconductor Science and Technology} \textbf{12} 1515-49

\bibitem{Holzmann}
Holzmann M, Tobben D, Abstreiter G, Wendel M, Lorenz H, Kotthaus J P and Schaffler F 1995 
\textit{Applied Physics Letters} \textbf{66} 833

\bibitem{Kouwenhoven}
Kouwenhoven L P, Marcus C M, McEuen P L, Tarucha S, Westervelt R M and Wingreen N S 
``Electron Transport in Quantum Dots" \textit{Mesoscopic Electron Transport} edited by L. L. Sohn, L. P. Kouwenhoven and G. Schon (Kluwer, Dordrecht, 1997) \textbf{345} 16-23

\end{thebibliography}
\end{document}